\input epsf
\documentstyle{article}
\voffset = - 1.7 cm
\hoffset = - 1.3 cm
\newcommand{\f}{\begin{equation}}
\newcommand{\ff}{\end{equation}}

\newcommand\tw{\tilde}

\begin{document}
\vspace{20 cm}
\rightline{\Large CGPG-94/2-1}
\rightline{\Large gr-qc/9402018}
\vfill
\centerline{\LARGE \bf Cosmological Histories  }
\centerline{\LARGE \bf  for the New Variables}
\vskip1cm
\centerline{Seth Major and Lee Smolin}
\vskip1cm
\centerline{\it   Center for Gravitational Physics and Geometry}
\centerline{\it The Pennsylvania State University}
 \centerline {\it University Park, Pennsylvania, 16802-6360   U.S.A.}
\centerline{February 7, 1994}
\vskip2cm
\centerline{ \bf Abstract}
\noindent
Histories and measures for
quantum cosmology are investigated through a quantization of the
Bianchi IX cosmology  using path integral techniques. 
The result, derived in the context of Ashtekar variables,
is compared with earlier work. A non-trivial correction to the measure 
is found, which may dominate the classical potential for universes on 
the Planck scale.   
\vfill
\eject

\section{Introduction}

Attempts to apply quantum mechanics to the universe 
have for some time divided into two main schools focused
 on either the canonical approach or path integral methods.  While these
 two approaches are driven by diffferent 
conceptions of how a quantum theory of the univererse is to
be constructed and interpreted, progress in these directions
has been circumscribed by the particular strengths and weaknesses
of the two formalisms.  These strengths and weaknesses
allow each formalism to address, in rather different ways,
key issues in quantum cosmology associated with the time
reparametrization invariance.  
This reflects a lack of lack of clocks and
observers ``outside'' the universe -- the  essential problem of quantum
cosmology. 

In the canonical formalism, the Dirac procedure gives us a precription
to construct physical states, define physical
observables and propose, through the imposition of the reality 
conditions of the theory, a physical
inner product.  The strength of the canonical approach 
arises in the way that
time reparametetrization invariance, as well as the other
gauge and diffeomorphism invariances of the theory can be
treated directly, yielding a gauge invaraint quantization.
The existance, for the full theory of quantum gravity, as
well as for models, such as $2+1$ gravity, one and
two killing field models, of exact results concerning physical
states and diffeomorphism invariant states and operators speak of
this strength.

On the other hand, the great weakness of the canonical approach
is that physical observables are very difficult to construct
explicitly because, both classically and quantum mechanically,
observables must commute with the Hamiltonian constraint and necessarily
freeze as constants of motion.  This difficulty is real; it reflects
the fact
that physical  operators which describe
time evolution ought to be constructed as correlations
between the degrees of freedom (one of which one would like to
take as a clock)\cite{carlo-time}.  For
example, suppose that we pick a condition that picks out a
slicing of spacetime into spacelike slices according to some
degrees of freedom of the theory.  Then we may define some
observables that measure spatially diffeomorphism invariant
information on these slices.  For example, 
let $A(q,p)$ be a spatially diffeomorphism invariant
quantity which measures an aspect of the geometry of
spacetime (where $q$ and $p$ are coordinates
on the phase space) and let $T(q,p)$ be another diffeomorphism
invariant quantity which we will take as measuring
time. Then for every possible value $\tau$ of this time
observable there is a {\it physical} observable
that measures what the value of $A(q,p)$ is 
on the spacial slice on which $T(q,p)=\tau$.   
We may note that as the condition that picks out the slices
is expressed in terms of the degrees of freedom this procedure
is completely gauge invariant because  within any gauge one can
specify these slices and evaluate the variables $A(q,p)$
and $T(q,p)$.  

While simple to specify, to express
such correlations explicitly in terms of functions on the phase space
or operators on the physical states one must solve the dynamics 
of the theory.  Thus, the construction of time 
reparametrization invariaint
observables in the canonical theory is a dynamical problem, which
one cannot expect to solve without approximation procedures for
theories outside of integrable and solvable systems.

Alternately, in this example using the path integral formalism
we find that the problem of taking the expectation values of
physical observables can be easily realized as soon as one has
a measure and a set of histories that represent physically
meaningful, gauge invariant amplitudes.   For
instance, the expectation value of the observable
$A(T=\tau )$ can be simply given by summing 
(with the appropriate measure)
paths weighted by the classical action and the value of the classical
observable $A(q,p)$ on the slice when $T(q,p) =\tau $,
\f
< \psi |\hat{A}(T=\tau )| \psi > = 
{ \int \left[ d\mu (q ,p )   A(T=\tau ) \right] e^{ { i \over \hbar } S} 
 \over 
\int \left[ d\mu (q ,p ) \right]  e^{{ i \over \hbar } S }  }.
\ff
 By varying $\tau$ we can describe the
evolution of
the system in terms of time reparameterization invariant quantities.
Though the path integral formalism steps by the difficulties of the
canonical approach, the path integral has complimentary difficulties.
Setting aside intepretational issues, we lack a prescription which
allows us to unambiguously find the set of histories, appropriate
contours, and a measure $\mu(q, p)$ which implement gauge invariances
and reality conditions.

We would like to suggest
that the situation points to
a mixed approach in which the 
physical, diffeomorphism invariant quantum states of the 
canonical theory are used as the starting point to define a
path integral and measure, after which the dynamics of physical
observables are computed with
path integral techniques.  
This program, if it can be concretely realized, offers
a possibility of an unambiguous formalism for quantum
cosmology making use of the strong points
of both the canonical and path integral approach.

To investigate this possiblity, it would be very useful
to have a working model of a quantum cosmology which has dynamics
complicated enough that problems of constructing
physical observables, inner product and path integral measure
are non-trivial.  However, the results ought to be simple enough that the path integrals for physically meaningful quantities
could be computed by relatively simple numerical or approximation
techniques.
This is the first of two papers which aim to 
lay the groundwork
to construct such a model of quantum cosmology
based on the Bianchi IX spatially homogeneous spacetimes. 
In this paper we show that a gauge invariant measure
can be constructed in this model, following
the Faddeev-Poppov procedure \cite{FAD} and  using the 
new variables \cite{newvar}.  In a companion
paper we consider a physically meaningful canonical
quantization procedure for Bianchi IX and show how,
and under what conditions,
quantities defined through this canonical formalism can
be expressed in terms of path integral expressions of the
kind that are derived here.

The  Bianchi IX
model describes a family of cosmologies in which
space is homogeneous, but the geometry  has two dynamical
degrees of freedom - measures of anisotropy.  It
has been studied extensively especially since 
the late 1960's when 
Misner found that, in a particular gauge, its dynamics
can be expressed in terms of the motion of a particle in a time-
dependent potential \cite{MTW}.  Though it is a simple system
with only two degrees of freedom, this model displays
a surprisingly rich behavior even at the classical level.   
For instance, it has been shown that the
Lyaponov exponent is greater than one for certain 
choices of time, meaning that the model is chaotic.\cite{chaos}
(However, the Lyapunov exponent, a measure of the exponential
separation of nearby trajectories in time, is 
not time reparameterization invariant!\cite{chaos,rugh}).
In the face of this it is unlikely that the theory can be exactly
solved, making it an ideal candidate to test the program
we have just discussed.

At the quantum level, although there does not exist,
to our knowledge, a complete
quantization of the Bianchi IX model
in either a path integral or canonical
formalism, a number of results
have been found previously.   An exact
physical state has
been found by Kodma \cite{KOD} using the new 
Hamiltonian variables of Ashtekar, which
can even be transformed into the metric representation
\cite{M&R}.  Graham has constructed a supersymmetric solution to
 this Bianchi model \cite{GRA}.  
Numerical work, following the
methods of Euclidean quantum cosmology\cite{BER} 
shows qualitative 
agreement with the exact solutions - at early 
times the wavefunction is spread over the aniostropy 
space while at later times the wavefunction peaks at 
the isotropic model (closed FRW).  

We find, perhaps not surprisingly, that the measure is 
non-trivial and dominates the weights of histories
when the radius of the cosomology is comparable to or
smaller than the Planck length.  This indicates that quantum 
effects dominate the behavior of the cosomology near the classical
singularity, as is generally expected.  
Using this result, we expect that it is now a
straightforward numerical problem to compute the expectation
value of physical observables in physical states.

We present the derivation in ``geometrized units'' in which
$G=c=1$.

\section{Bianchi IX in the new variables}

The new variables provide a complex chart on the phase space of general 
relativity with configuration variables, the 
connections $A_a^I$, and conjugate momenta, the densities
$\tw{E}^a_I$.  Our notational convention denotes spatial indices as
lower case latin letters, e.g. 
$ a, b, c, ...$ and denotes internal indices as  upper case latin letters, e.g. I, J,....  Densities of weight one, such as the conjugate momenta, $\tw{E}^a_I$, claim a tidle.  The phase space is endowed with the structive given by
\f
\{ A_a^I (x) , \tw{E}^b_J (y) \} = i \, \delta_a^b \, \delta^I_J \,
\delta^3(x,y).
\ff
The more familiar metric is obtained 
from the frame fields, $\tw{E}^a_I$, by defining triads on a three-
manifold $\Sigma$, $E^I_i = {1 \over \sqrt{h}} \tw{E}^I_i$, and
by letting $h^{ij} = E^I_i E_{Ij}$. As this  chart is a complex
one, to regain general relativity we must 
choose a  section of the phase space in
which reality conditions, such as

\f
\left ( {h_{ij}} \right )^* = h_{ij} \label{realitycond1}
\ff
and
\f
\left ( {\dot{h_{ij}}} \right )^*= h_{ij}  \label{realitycond2}
\ff
hold.

In the 3+1 decomposition, with 
$\sigma$
chosen as the time parameter, the classical action is
\f
I[A,E,N] = \int_{\sigma_i}^{\sigma_f} d\sigma \int_\Sigma d^3x 
\left (-i \tw{E}^I_a \dot{A}_I^a - N_* C^* \right ).
\ff
The asterisk, *, is an index that runs 
from 0 to 6; it has one value for each constraint:
\f
S := \varepsilon_{IJK} F^I_{ab} \tw{E}^{aJ} \tw{E}^{bK} = 0 \label{sconst}
\ff
\f
G_I := D_a \tw{E}^a_I = 0 \label{gconst}
\ff
\f
V_a := F_{ab}^I \tw{E}_I^b = 0 \label{vconst}
\ff
which are known as the scalar 
or hamiltonian, gauss, and vector or diffeomorphism constraints,
 respectively.  
The covariant derivative is associated with the
connection $
D_a f^I := \partial_a f^I +\varepsilon^{IJK} A_{aJ} f_K $
and the curvature,
$ F_{ab}^I := \partial_{[a} A_{b]}^I + 
\varepsilon^{IJK} A_{aJ} A_{bK}.$
We invesitage Class A Bianchi IX models.~\footnote{The classification of 
Bianchi
models involves splitting the structure constants of the Lie group of
isometries  into two irreducible pieces.  Denoting these by $S^{LI}$ and 
$V_K$,
the  strcuture constants may be written as 
$C^I_{JK} = \varepsilon_{JKL} S^{LI} + \delta^I_{[J} V_{K]}$.  
Class A models are those for which 
$V_I = 
0$.}  Homogeneity provides us with a 
foliation of spacetime into homogeneous space-
like surfaces and gives each leaf a 
left inavriant vector- one-form basis, $(v,\omega)$ in which to 
expand 
the new variables \cite{nenad2}.  On each leaf we can write
\f
A_a^I = a_S^I(\sigma ) \omega_a^S (x) \label{as}
\ff
and
\f
E^a_I = e^S_I (\sigma ) v_S^a (x) \label{es}.
\ff
These expansion coefficients may be viewed as $3\times3$ 
matrices.  Homogeneity merrily reduces field theory to mechanics - from $9\times9$ degrees of 
freedom per spacetime event to
$9\times9$ for each spatial section.  The action  
simplifies to:
\f
I[A,E,N] = \int_{\sigma_i}^{\sigma_f} d\sigma 
\left (-i \Omega e^I_L \dot{a}_I^L - N_* C^* \right ). \label{action}
\ff
Here, $\Omega = \int_\Sigma \omega \wedge \omega \wedge \omega
 = 16\pi^2$ is the volume
 element on $SU(2)$. (The lagrange 
multipliers have been rescaled.)  Henceforth, we will work in the unusual
units $G=c=\Omega=1$ meaning we measure
fields in terms of this volume element and use a  conversion factor of
$c^4/ G (\Omega)^{1/3}$ for energy.  In terms of
 the expansion 
coefficients defined in Eqs. (\ref{as}) and (\ref{es}), 
the constraints, Eqs. (\ref{sconst}), (\ref{gconst}), and (\ref{vconst}) become
\f
S = \varepsilon_A^{BC} \left ( -\varepsilon^D_{GF} a_D^A +
\varepsilon^A_{DE} a^D_G a^E_F \right ) e^G_B e^F_C
\label{unfixedHconstraint}
\ff
\f
G_I = \varepsilon_{IJ}^K a_L^J e^L_K 
\ff and 
\f
V_J = \varepsilon^I_{JK} a^L_I e^K_L.
\ff
We choose to fix the six gauge and diffeomorphism constraints
by a "diagonal gauge" \cite{nenad2} to yield the extensively studied form of
Misner in which the 
cosmology may be seen as a particle moving
in $2+1$ dimensional spacetime with a time dependent potential.
This choice parallels Misner's $\beta_+, \beta_-$ 
diagonalization in the geometrodynamic framework (we
will later translate our result into Misner's notation for comparison).

We define
\f
\epsilon_1 := e^1_1, \epsilon_2 := e^2_2,  \epsilon_3 := e^3_3
\ff
and choose
\f
\chi^I_J \equiv e^I_J=0   \mbox{  for }  I \neq J. \label{gauge}
\ff
Upon imposing these conditions the three Gauss contsraints vanish, while the three remaining vector constraints require that the off-diagonal
components of $a^I_J$  vanish as well.  As above we define
\f
a_1 :=a^1_1, a_2 := a^2_2,  a_3 := a^3_3.
\ff
At the end of the kinematical
gauge fixing, we are left with six 
canonical degrees of freedom per leaf.  

At this kinematical level in which the Gauss and vector
constraints have been solved, but the Hamiltonian constraint
have not, the model is not difficult
to quantize.  The six canonical degrees of freedom
can be taken to be diagonal components of the frame
fields and (imaginary parts of) diagonal components of the
connections.  States, in the diagonal metric
representation, may be expressed as functions
of the $\epsilon$'s.  The reality conditions, Eq. (\ref{realitycond1})
and Eq. (\ref{realitycond2}), are realized by the inner product,
\f
\left< \psi(\epsilon)| \phi(\epsilon ) \right> = \int d^3 \epsilon \,
e^{-F(\epsilon)} \bar{\psi(\epsilon)} \phi(\epsilon). \label{innerproduct}
\ff
where
\f
F(\epsilon) :={\epsilon_1 \epsilon_2 \over \epsilon_3} +
{\epsilon_2 \epsilon_3 \over \epsilon_1} +
{\epsilon_3 \epsilon_1 \over \epsilon_2} . \label{fdef}
\ff
Unfortunately this quantization cannot be used to compute
physical quantities; reparametrization invariance remains.  The
the Hamiltonian constraint must be solved to reduce
 the physical phase space to four degrees of freedom.

To pull out the dynamics, we fix this reparameterization 
invariance by choosing the gauge condition:
\f
\chi^0 (\sigma ):= \ln(\sqrt{h}) - \sigma=0 \label{timechoice}
\ff
fixing $\sigma$ to be proportional to the volume of each spatial
surface.   This choice of paramterization is monotonic on half the
history of any given classical Bianchi IX cosmology\cite{rugh}
(Bianchi IX expands from an initial singularity and then 
collapses in a final singularity \cite{LIN}).
An
alternative choice exists which is monotonic on the whole of the
evolution.  This is to
choose the time $\sigma$ to be proportional to the 
momentum
conjugate to the $\ln(\sqrt{h})$ -- the
trace of the extrinsic curvature of the slices of
homogeneity.   We have not studied the form
of the path integral in this gauge in detail, but see no apparent
reason why this could not be carried out.

To proceed to completely specify the canonical quantization
in this gauge we should find a complete set of 
physical coordinates
and momenta on the 
subspaces labeled by $\sigma$.   We do
not know how to do this. Fortunately, the facillity of the Faddeev-
Popov ansatz allows us to compute the path integral.

\section{Construction of the path integral}

Ideally, we would provide a chart for the physical phase
space, use this chart as the groundwork of a operator algebra,
endow the space of states with an inner product, and produce
dynamics through a hamiltonian composed of these operators.  We  denote the
 cannonical coordinates as
$\bar{q}^\iota (\sigma) )$ and  
$ \bar{p}_\iota (\sigma)$ (where $\iota=1,2$) and denote the set 
of eigenstaes by
$|\bar{q}^\iota (\sigma )>$ and 
$ |\bar{p}_\iota (\sigma )>$.  As 
classically
$\bar{{}q}^\iota (\sigma )$ and 
$\bar{p}_\iota (\sigma )$ are canonically conjugate, we would have
$\left< \bar{q}^\iota (\sigma )|\bar{p}_I (\sigma \right>
=\exp( i \bar{q}^\iota \bar{p}_\iota) (\sigma ).$
If the hamiltonian
which realizes evolution from 
fixed volume slice to fixed volume slice is $h( (\sigma ))$, then we would
have,
\begin{eqnarray}
<{q}^\iota_f (\sigma_f )| {q}^\iota_i 
(\sigma_i) >_{phys}
&=& K \left ( {q}^\iota, \sigma_f ; {q}^{\iota} , 
\sigma_i \right )
\nonumber \\
&=&
\int \left [ d\bar{p} \, d\bar{q}  \right  ] \exp{ \, i \int 
\left (
\bar{p} \dot{\bar{q}} - h(\bar{p}, \bar{q}) \right ) dt}. 
\label{phys-prop}
\end{eqnarray}
where the integral is over all possible physical trajectories which
pass through the initial point  $q^\iota_i$, at volume
$\sigma_i$ and the final point 
$q^\iota_f $, at volume
$\sigma_f$. The brackets indictate that the measure
is taken on each time slice of the skeletonization of the path integral.
The brackets also include a factor of $1/\sqrt{2\pi}$ for each differential.
This notation will be used for the remainder of this paper.

This construction is only useful through its link to
 a path integral over the whole (unphysical)
phase space.  Denoting the coordinates and momenta of
the whole phase space by $q$ and $p$ \cite{FAD},
\begin{eqnarray}
K \left ( {q}^\iota_f , \sigma_f ; {q}^\iota_i , \sigma_i \right )
& = &
\int \left [ {dp \, dq \over 2 \pi} {dN \over 2 \pi} \prod_* 
\delta(\chi^*)  \left| \{ C, \chi \} \right| \right ] \nonumber \\
& & \times \exp{ i \, \int \left (  p
\dot{q} - h(p, q) - N^*C_*(p, q) \right ) dt}  \label{ungauge-prop}
\end{eqnarray}
where the $C$ are the constraints of the theory, the $\chi^*(p,q)$ are the
gauge choices.  The key element of the link, the determinant, involves a
Poisson bracket between constraints and gauge fixing conditions.  The time coordinate $\sigma$ is an arbitrary parametrization of the phase space trajectories and the initial and final conditions of the path integral
are chosen to agree with those in Eq. (\ref{phys-prop}). ( The standard procedure for
gauge theories described in \cite{FAD} generalizes
to the case in which time reparameterization
invariance is one of the gauge invariances, so long as
the choice is consistant\cite{nenad1}.)

In our example of Binachi IX, the kinematical gauge symmetries are fixed
by the choice of a diagonal gauge and the time reparametrization
invariance is fixed by associating time with volume.  Writing the physical
 coordinates ($\bar{q}$ and $\bar{p}$ above) as the anisotropies, $\beta$, a
 transition element from $\sigma_i$ to $\sigma_f$ is generated by
 integrating the intial state with the kernel,
\f
\left< \beta | \psi(\sigma_f) \right> 
= \int \left [ d \beta \right ]
K \left ( \beta, \sigma_f;
\beta, \sigma_i \right )
\left< \beta | \psi(\sigma_i) \right>.
\ff
The kernal, the object we shall be concerned with
from now on, may be written in the new variables as
\f
K (\epsilon_f, \sigma_f ;\epsilon_i, \sigma_i) = 
\int
\left [ d^3 \epsilon \, d^3a \, 
\left| \{ C^* , \chi_* \} \right|
\delta [\chi^0(\epsilon, \sigma)] \delta(S) \right]
\exp i \int_{\sigma_i}^{\sigma_f} d\sigma 
\left( -i \epsilon^T  \dot{a} \right)  .
\label{prop.5}
\ff
Exponentiating the measure's delta-function as
\f
\delta(S) = \int_{-\infty}^{\infty} \delta_\sigma
 dN e^{ - i \delta_\sigma N S} ,   \label{expoN}
\ff
in which $\delta_\sigma$ is the step size  of the skeletonization
of the path integral, we can write the kernal of Eq. (\ref{prop.5}) as
\f
K (\epsilon_f, \sigma_f ; \epsilon_i, \sigma_i) =
\int
\left[ d^3 \epsilon \, d^3a \, \delta_\sigma dN \, 
\left| \{ C^* , \chi_* \} \right|
\delta [\chi^0(\epsilon, \sigma)] \right] 
\exp{  i \int_{\sigma_i}^{\sigma_f} d\sigma 
\left (-i \epsilon^T  \dot{a} - N S \right ) }.
\label{prop1}
\ff
with the action in the form of the (gauge-fixed) action of Eq. (\ref{action}).
In Eqs. (\ref{prop.5}) and (\ref{prop2}) we performed the trivial integration over the off-diagonal pieces of $a^{I}_{J}$ and $e_{I}^{J}$, eliminating  vector and diffeomorphism constraints and their
assoicated gauge section $\delta $-functions.  However, the measure contains contributions from both the kinematical gauge 
fixing and the time reparametrization gauge
fixing (which is still explicitly indicated).  The effects of our gauge choices
can be computed explicitly,
\begin{eqnarray}
\left| \{ C^*, \chi_* \} \right| &=&   \left| \begin{array} {ccccccc}
0 & 0 & 0 & 0 & -\epsilon_2 & 0 & \epsilon_3  \\
0 & 0 & \epsilon_1 & 0 & 0 & -\epsilon_3 & 0  \\
0 & -\epsilon_1 & 0 & \epsilon_2 & 0 & 0 & 0 \\
0 & 0 & 0 & 0 & \epsilon_3 & 0 & -\epsilon_2  \\
0 & 0 & -\epsilon_3 & 0 & 0 & \epsilon_1 & 0  \\
0 & \epsilon_2 & 0 & -\epsilon_1 & 0 & 0 & 0 \\
F(\epsilon) & 0 & 0 & 0 & 0 & 0 & 0 
\end{array} \right|
\nonumber \\
&=& \left| {\epsilon_2^3 \epsilon_3^3
\over \epsilon_1}  \left (\epsilon_3^2 - 
\epsilon_2^2 \right ) 
+
{\epsilon_1^3 \epsilon_3^3
\over \epsilon_2} \left (\epsilon_1^2 - \epsilon_3^2 \right ) 
+
{\epsilon_1^3 \epsilon_2^3
\over \epsilon_3} \left (\epsilon_2^2 - \epsilon_1^2 \right ) \right|
\end{eqnarray}
where $F(\epsilon)$ is defined in Eq. (\ref{fdef}).  As the Hamiltonian constraint in Eq. (\ref{unfixedHconstraint}) contains terms 
both linear and quadratic in the $a^I_J$'s and as the gauge fixing
condition is a function of the $e^I_J$'s, it is possible that terms 
linear in the $a^I_J$'s appear in $\left| \{C^*,\chi_*\} \right|$.
However as a result of a serendipitous simplification, such terms cancel. 
The remaining constraint -- the hamiltonian constraint -- is written
\f
S = \left ( - a_1 + a_2 a_3 \right ) \epsilon_2 
\epsilon_3 +
\left ( - a_2 + a_3 a_1 \right ) \epsilon_3 \epsilon_1 
+
\left ( - a_3 + a_1 a_2 \right ) \epsilon_1 \epsilon_2.
\ff
The imposition of this constraint is enforced by the integral
over $N$, which has a range from 
$-\infty$ to $+ \infty$.  The parameterization, Eq. (\ref{timechoice}), restricts the range of the $\epsilon$'s integration to the
positive real axis $(0, +\infty )$.  Meanwhile, restricting the $\epsilon$ integral to the real axis
satisfies the reality conditions of Eq. (\ref{realitycond1}) that require that
the three metric be real.

To implement the rest of the
 reality conditions, Eq. (\ref{realitycond2}), 
we may chose a contour
for the $a_I$ integral that reflects these conditions.
Recall that  the $A_a^I$'s are complex
variables which depend on the original canonical
variables of relativity via 
$ A_a^I = \Gamma_a^I (E) + i K_a^I (E, \Pi) $
where $\Gamma_a^I$ is the $SU(2)$ connection, $K_a^I$ is the 
extrinsic 
curvature, and $\Pi$ is the canonically, conjugate momentum to $E$.
We have a similar relation for the diagonalized expansion 
coefficients,
\f
a_I = \gamma_I (\epsilon) - i \kappa_I.
\ff
The reality conditions suggest that the $\kappa_I$'s may be taken
as the independent variables in the path integral.  The
countours in Eq. (\ref{prop1}) may then be taken along the
imaginary $a$ axes or, equilvalently, 
performed for real $\kappa$.  At each $\sigma$
\f
\int d^3\epsilon d^3 \kappa = \int d^3 \epsilon d^3 a.
\ff
After integrating by parts and discarding a complex boundary term 
($ - \epsilon^T \kappa |_{\sigma_f}+\epsilon^T \kappa |_{\sigma_i}$)  the propagator of Eq. (\ref{prop1}) becomes 
\begin{eqnarray}
K (\epsilon_f, \sigma_f ; \epsilon_i, \sigma_i) &=&  
\int_0^{\infty} \left [ d^3 \epsilon \right ] \int_{-\infty}^{\infty} 
\left [ d^3 
\kappa \right ] \int_{-\infty}^{\infty} 
\left [ \delta_\sigma dN 
\left| \{ C^*, \chi_* \} \right|
 \delta [\chi^0 (\epsilon, \sigma)] \right ]
\nonumber \\
&\times &
\exp  i  \left ( \int_{\sigma_i}^{\sigma_f} d \sigma \kappa^T Q \kappa 
+ b^T \kappa \right )    \label{prop1.5}
\end{eqnarray}
where the coefficient of the quadratic term is
\f
Q = { N \over 2} \left ( \begin{array} {ccc}
0 & \epsilon_1 \epsilon_2 & \epsilon_1 \epsilon_3 \\
\epsilon_1 \epsilon_2 & 0 & \epsilon_2 \epsilon_3 \\
\epsilon_1 \epsilon_3 & \epsilon_2 \epsilon_3 & 0 
\end{array}
\right ) 
\ff
and the coefficient of the linear term is
\f
b =  \left ( \begin{array}{c} \dot{\epsilon_1} \\ \dot{\epsilon_2} \\ 
\dot{\epsilon_3} \end{array} \right )
- i N \left ( \begin{array}{c} \epsilon_2 \epsilon_3 \\ 
\epsilon_1 \epsilon_3 \\ \epsilon_1 \epsilon_2 \end{array} \right ).
\ff
The integration over $\kappa$  may be
done; it is gaussian.  This integral exits when the matrix $Q$
is a real, positive matrix which is ensured by the reality conditions
(Eq. (\ref{realitycond1})) and the choice of parametrization, respectively.
The cross terms generated by the integration form a total derivative of the
measure factor in the unphysical inner product, Eq. (\ref{innerproduct}).
Therefore, the 
integration preserves the configuration space -  to -  configuaration space
propagator
under this inner product.

Performing the integration over  
$\kappa$ then gives,    
\f
K (\epsilon_f, \sigma_f ; \epsilon_i, \sigma_i) = 
\int \left [ d^3 \epsilon \, 
{\sqrt{\delta_\sigma} dN \over \sqrt{N}} 
\, \mu  (\epsilon) \, 
\delta [\chi^0 (\epsilon, \sigma)] \right ]
\exp - i \int_{\sigma_i}^{\sigma_f} d\sigma_N 
L(\epsilon, \dot{\epsilon}, N )
\label{prop2}
\ff
with the measure; 
\begin{eqnarray}
{\mu} (\epsilon)  & =&
% \nonumber \\
   \left| \left ( \epsilon_2
\epsilon_3 \over \epsilon_1 \right )^2 \left (\epsilon_3^2 - 
\epsilon_2^2 \right
) + \left ( \epsilon_1 \epsilon_3
\over \epsilon_2 \right )^2 \left (\epsilon_1^2 - \epsilon_3^2 \right ) 
+
\left ( \epsilon_1 \epsilon_2
\over \epsilon_3 \right )^2 
\left (\epsilon_2^2 - \epsilon_1^2 \right ) \right|
\end{eqnarray}
and the lagrangian;
\begin{eqnarray}
L(\epsilon, \dot{\epsilon}, N) &=& { 2 \over N} 
\left [ \left ( {\dot{\epsilon_1} \over
\epsilon_1} + {\dot{\epsilon_2} \over \epsilon_2} 
+{\dot{\epsilon_3}
\over \epsilon_3} \right )^2 - 4 \left (
{\dot{\epsilon_1} \dot{\epsilon_2} \over \epsilon_1 \epsilon_2} +
{\dot{\epsilon_2} \dot{\epsilon_3} \over \epsilon_2 \epsilon_3} +
{\dot{\epsilon_3} \dot{\epsilon_1} \over \epsilon_3 \epsilon_1}
\right ) \right ]  \nonumber
\\
&&- 2 N \left [ F(\epsilon)^2 - 4 \left ( \epsilon_1^2 + 
\epsilon_2^2 + \epsilon_3^2 
\right ) \right ]
\end{eqnarray}
The integral (\ref{prop2}) is the main result of our analysis.  
It is manifest 
that the dynamics unfolds with a time-dependant potential though, 
the form of
the potential is unclear.  A more physical picture may be found by
 re-expressing our result in terms of anisotropy.
To accomplish this and to compare with previous studies of the
 Bianchi cosmology
we translate Eq. (\ref{prop2}) 
back into Misner's chart on the phase space of
Bianchi IX using
\f
h_{ij} = e^{2 \alpha} \left ( e^{2\beta} \right )_{ij} =  
\epsilon^2_i \delta_{ij}
\ff
where $\beta$ is the traceless matrix: 
diag$\left ( \beta_+ + \sqrt{3}\beta_-,\beta_+ - \sqrt{3}\beta_-, 
-2\beta_+ \right )$.
The Jacobian of this transformation from the $\epsilon$ chart into the
$\beta_\pm$ chart is
\f
\left| { \partial \epsilon_1\epsilon_2\epsilon_3 \over \partial
\beta_\pm \partial \alpha } \right| =  6 \sqrt{3} e^{3 \alpha}.
\ff
The integral of Eq. (\ref{prop2}) may then be written
\begin{eqnarray}
K (\beta_f, \sigma_f ; \beta_i, \sigma_i) & = & \int_{-\infty}^{\infty}
\left[ d\beta_+ d\beta_- d\alpha 6\sqrt{3} {dN \sqrt{\delta_{\sigma} 
\over N} } \delta( \sigma - 3 \alpha)
e^{7\alpha} \mu(\beta_\pm) \right] \nonumber \\
& & \times \exp 2 i \int d\sigma_N { 1 \over N }  
\left( 12 \dot{\beta_+}^2 + 12 \dot{\beta_-}^2 - 3 \dot{\alpha}^2 \right)
-N e^{2\alpha} U(\beta_\pm)
\end{eqnarray}
where $\mu(\epsilon)=e^{-4\alpha}\mu(\beta_\pm)$ and
\f
\mu(\beta_\pm)=| e^{8 \beta_+}
\sinh ( 4 \sqrt{3} \beta_- )
+ e^{-10 \beta_+} \sinh( 2 \sqrt{3} \beta_- ) - e^{2\beta_+} 
\sinh (6 \sqrt{3}\beta_-)|.
\ff
Letting the delta function consume the $\alpha$ integration we find
\begin{eqnarray}
K (\beta_f, \sigma_f ; \beta_i, \sigma_i) &=& \int \left [ d\beta_+
d\beta_- { dN \over \sqrt{N}} 6 \sqrt{3\delta_\sigma}
e^{7/3 \delta_\sigma}  \mu(\beta_\pm)
\right ]
\nonumber \\
&& \times \exp 2 i \int_{\sigma_i}^{\sigma_f} d\sigma_N 
{1 \over N} \left( 12 \dot{\beta_+}^2 + 12 
\dot{\beta_-}^2 -{1 \over 3} \right)
 - N e^{{ 2 \sigma \over 3}} U(\beta_\pm). \label{betaprop}
\end{eqnarray}

The propagator's form reveals its character.
The action is that of Misner's Bianchi IX formulation of a particle
moving in a time  dependent potential while $N$ has two roles,
 evolving the theory it
and preserving reparameterization invariance as it does for 
the relativistic particle\cite{relpartpi}.
\begin{figure} [h]
\epsffile{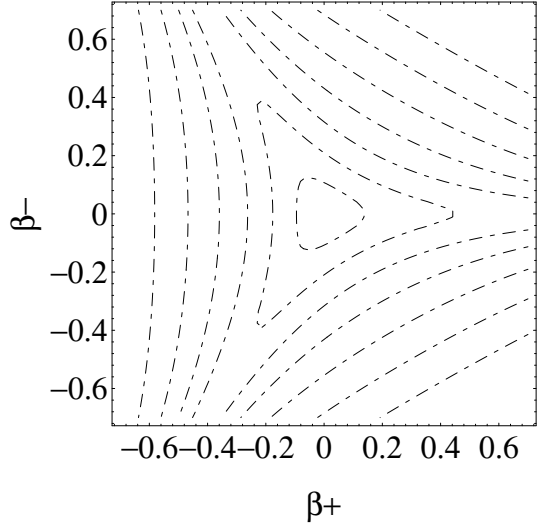}
\caption{The classical potential for Bianch IX with contours spaced
by powers of e with the first line at $U(\beta_\pm)=-0.9$. Three 
``channels,'' 
one along the positive $\beta_-$ axis, and the other two 
sloping diagonally
in the negagative $\beta_-$ axis indicate the minima of the potential. 
 The  contours form a rough triangle with steep walls.}
\end{figure}
The action contains the classical potential,
$V(\beta_\pm ) := U(\beta_\pm) +1$ 
which takes the usual form\cite{MTW}
\footnote{Note that in
the literature one often finds written 
$V(\beta_\pm ) := U(\beta_\pm) +1$.  This is convenient because
$V(\beta_\pm)$ is bounded from below, however what is
important to remember is that the actual potential
$U(\beta_\pm$) is bounded from below by $-1$.}
\begin{eqnarray}
U(\beta) &= &{1 \over 3} e^{-8 \beta_+} 
- {4 \over 3} e^{-2 \beta_+}
\cosh (2 \sqrt{3} \beta_-) 
\nonumber \\
&& +
{2 \over 3} e^{4\beta_+} \left (\cosh (4 \sqrt{3} \beta_-) -1 \right ).
\end{eqnarray}
A contour plot of this potential appears in Figure 1.
The evolution of the
cosomology is seen as dynamics of a free particle reflecting off
roughly triangular, expontially steep ``walls'' shown in Fig. 1.  
 
The nontrivial measure shows the mechanics of the gauge fixing 
proceedure and
the traces of integration. A weighting factor, constant in $\beta_\pm$
but proportional to $1/N$ appears in the action.  Through the
process of skeletonization, this weight may be expressed in the
measure (on each time slice) as $\exp \left( {2 \delta_\sigma \over 3N}
\right)$.  The $\beta$-dependent form of $\mu(\beta_\pm)$,
 depicted in Figure 2,
\begin{figure} [h]
\epsffile{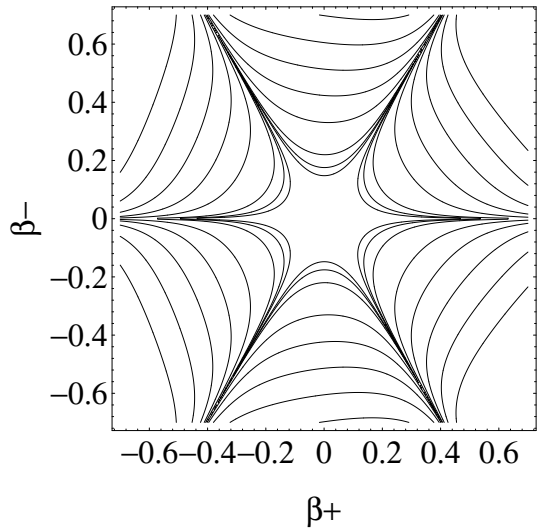}
\caption{A plot of the measure of the path integral for Bianchi IX. 
The contours are separated by powers of e with the first contour 
at $\mu(\beta_\pm)= 0.468$.The six channels show where histories 
are unsupported by the measure.}
\end{figure}
has a six-fold symmetery.  The points of the point are the minima and so 
 lend little support for
 wavefunctions peaked in these regions.  In particular, 
the measure tends to
support histories which do not peak in the channels of the 
classical potential
and do not have maxima near the center of the triangular walls.  
Therefore,
it seems that the effect of the measure is to splinter a 
wavefunction near the walls.

The measure factor, $\mu(\beta_\pm)$,
may be expressed as
as a ``quantum correction'' to the classical lagrangian by
exponentiating it giving a ``quantum lagrangian'' (including $\hbar$),
\f
L_q(\beta_\pm, \dot{\beta_\pm}, N ) = {12 \over N} 
\left( \dot{\beta_-^2} + \dot{\beta_+^2} \right)
-N e^{2\sigma \over 3} U(\beta_\pm) - i \hbar V_q(\beta_\pm)
\ff
with the measure  $\mu(\beta_\pm)$ expressed in terms of a quantum
correction $V_q$ to the potential as
\f
V_q( \beta_\pm) = \ln [ \mu(\beta_\pm ) ]
\ff

The expression of Eq. (\ref{betaprop})
 is as far as we believe
can be gone in the evaluation of the path integral of the
Bianchi IX cosmology, without turning to approximation or
numerical techniques.  We may note that to evaluate either of these
propagators, Eq.(\ref{betaprop}) or Eq. (\ref{prop2}),
it will be necessary to perform a Wick rotation. Since the ``lapse,'' $N$, 
is
 used to exponentiate a
delta function, the range of integration runs from $-\infty$ to $+\infty$.
However, the construction requires us to include only forward
(in the time parameter $\sigma$) propagating histories.  To secure 
convergence
in the anisotrpy propagator Eq. (\ref{betaprop}) and in the quadratic
integration Eq. (\ref{prop1.5})
we must rotate to 
$\sigma \rightarrow \sigma_N$, where
$\sigma_N=\sigma \exp \left( i {\pi \over 2} \right)$
when $N >0$ and 
$\sigma \rightarrow \sigma_N = \sigma \exp \left(- i {\pi \over 2} \right)$
when $N< 0$, effectively excluding ``backwards evolving'' histories.
Alternately, it's possible to restrict
the $N$ integration at the onset by using a $\theta$-function in the 
exponentiation, Eq.
(\ref{expoN}), as may be done in the path integral for the relativistic
particle \cite{relpartpi}.  It is reassuring that the 
$N$ dependent
rotations necessary to make the $\alpha$ integration convergent
will serve again to make the $\beta_\pm$ integrals
convergent.   In particular, 
if the time is rotated to purely imaginary values
we see that the classical part of the Euclideanized action is
indeed positive definite. Moreover, the measure factor is
purely real and positive, so that its presence does not disrupt
this issue.  Thus, as expected from general
arguments\cite{emilpawel}, when the path integral is defined
through a correct gauge fixing procedure there is no problem
with the "runaway conformal modes" of the Euclideanized action.
The analytic continuation provides a context to compare the two
potentials as shown in Figure 3.
\begin{figure} [h]
\epsffile{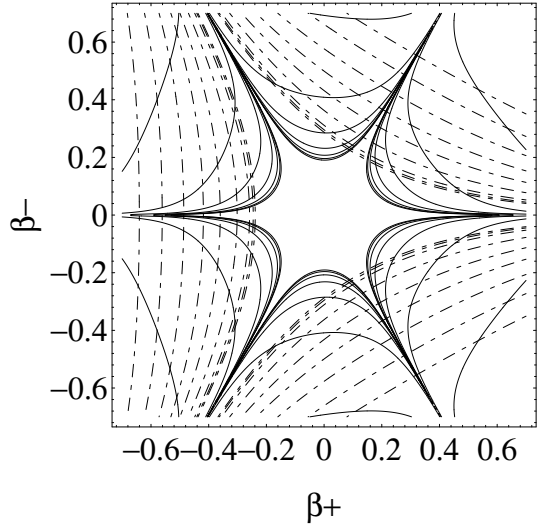}
\caption{The classical potential and ``quantum potential''
superimposed for direct comparison.  The effect of the measure is 
to suppress amplitdues in the three channels and at the center of 
the three walls of the classical potential. The contours (of both
potentials) are spaced by multiples of two beginning at 0.1 (e.g.
the first contour is drawn when the potentials have value 0.1, 
the second
contour at 0.2, the third at 0.4, etc.) }
\end{figure}
For a fixed
value of the classical potential, as the cosmology approaches the
 initial singularity ($\sigma
\rightarrow -\infty)$ the effect of the classical potental vanshes while
the ``quantum potential'' remains.  In fact this suppression occurs for any
interval during the first half of the history of the cosmology (for
$\sigma<0$).

Unfortunately, there are  drawbacks to analytically continuing
the paramter $\sigma$ as may be seen in quantum action:
\f
I_q(\beta_\pm, N)= -\int d\sigma {2 \over N} \left( 12 \dot{\beta_+}^2
+ 12 \dot{\beta_-}^2  + {1 \over 3} \right) + 2N e^{i {2 \sigma \over 3} }
U(\beta_\pm) - V_q(\beta_\pm)
\ff
in which $\sigma \rightarrow i\sigma$ and so all variables displayed
 are real.  As the
 potential term $Ne^{2\sigma/3}U(\beta_\pm)$ contains a
paramter dependent factor, when we continue to imaginary time the potential
aquires an unfortunate oscillation. However, this may yield the expected
result for as we go 
towards the intial singularity ($\sigma \rightarrow -\infty$),
a small change of $\sigma$ near the initial singularity causes this
factor to oscilate wildly, effectively averaging the potential to zero.  
It is
unclear whether the dynamics is correctly modeled by this continuation of
$\sigma$.

But, there is another choice of analytic continuation.  The analytic 
continuation of $\sigma$ may be seen as continuing the determinant of the
metric -- the $e_I$'s -- which breaks our reality conditions. To 
avoid the oscillating potential and this problem we
could instead continue $N$.  This produces a positive classical
lagrangian given by
\f
I_q(\beta_\pm, N)= -\int d\sigma {2 \over N} \left( 12 \dot{\beta_+}^2
+ 12 \dot{\beta_-}^2 - {1 \over 3} \right) + 2N e^{i {2 \sigma \over 3}}
U(\beta_\pm) - V_q(\beta_\pm)
\ff
where all the displayed quantities are real.  The form of the action
is identical expect one sign change; the weight factor proportional
to $1/N$ changes sign.
Since $g_{00}= -N^2$ this continuation would give
us a Euclidean metric.   
A possible intepretation of this weight is that it gives weight to free
particle histories or ``Kasner'' epochs.  
To see this we should return to the definition 
of the path integral in terms of skeletonized paths.  
In the limit of small, $N<<1$, 
for any fixed value of $\beta_\pm$, the 
factor $\exp { 2 \delta_{\sigma} \over 3 N}$ gives 
weight to the kinetic
energy term while the potential is negligable; this factor 
tends to support free particle histories.  With the opposite sign
in the $\sigma$ analytic continuation, the free particle histories
are suppressed.  This effect is not large (for the
path integral is defined in the limit  
$\delta_{\sigma_N} \rightarrow 0$) but
nevertheless the action with $N>>1$ reduces to the free particle.
Until physical quantities are computed it is unclear precisely how these
continuations are related.

\section{Conclusion}

We used the Faddeev-Poppov prescription to construct
a path integral for the Bianchi IX quantum cosmology.  Our
 strategy is to begin with a classical dynamical
system on a $9+9$ dimensional complex
phase space
defined by the constraints ( Eqs (\ref{sconst}), (\ref{vconst}),
 and (\ref{gconst})) and define the measure of
the path integral which follows from
the Faddeev-Poppov ansatz.  We choose contours of integration which
ensure that phase space histories
satisfy the reality conditions, corresponding to the 
physical condition that the metric of
spacetime is real and of Minkowskian signature.  

We find, as a result,  the configuaration space -- anisotropy 
space -- path
integral Eq. (\ref{betaprop}) for the Bianchi IX cosmology in 
the gauge in which time is parameterized by the
volume of space.  RQuantum correctionsS
to the potential arise in performing gauge fixing with the
Faddeev-Poppov method yielding the non-trivial effects
of the measure.
We may make several comments about its form.

First of all, it is interesting to note that the effect of 
the measure is
independent of time, and hence the volume of the universe.
However, the relative importance of the classical and
quantum parts of the potential are time dependent
as the classical potential is multiplied by 
the factor $e^{2 \sigma /3}$.  
By graphing $\mu (\beta_\pm )$, as in Fig. 1,
 for a fixed $\sigma$ we
may see where the effects of the measure are important.

Comparing the quantum and classical
potentials (as we do in Fig. 3) the quantum effects are negligable for
large anisotropies except for times much less than 
the Planck time.  Furthermore, 
relative region of the $\beta_+, \beta_-$ plane where
the quantum effects dominate grows smaller
as time increase because of the factor of 
$e^{\sigma}$ in the classical potential.  Of course, 
the fact that the quantum potential diverges
for vanishing anisotropy merely 
reflects the fact that the measure vanishes, when the
anisotropy $\beta_-$ vanishes.  This means that the 
$\beta_-$ - isotropic evolutions have vanishing measure for all time, 
which implies that the quantum state must have
a finite spread in anisotropy.   We may note that this effect
is stronger than a simple uncertainty principle effect in that
it might be expected to keep the "wave function of the 
universe" spread in anisotropy.
The measure actually vanishes when
$\beta_-=0$, which is  the configuration space of the Taub model. 
This suggests
that reductions from one quantum cosmological
model to another are not always appropriate.

To further understand the effects of the measure in the
path integral, it is necessary to  finish the evaluation of the
path integral.  As we have seen that there is an analytic continuation
which makes the integral in Eq. (\ref{betaprop}) real and
convergent, there should be no problem with defining
the integral through standard Monte-Carlo techniques,
or by semiclassical techniques.

In particular, given this kernal  one can proceed directly to the
evaluation of 
the expectation values of
gauge invariaint, and hence physically meaningful, quantities.  
For example, any quantity of the form 
$F(\beta_\pm, (\sigma ))$ measures correlations of the
anisotropies, defined at slices with particular volumes,
is gauge invariant and meaningful\cite{carlo-time}.
Quantities like this have been evaluated successfully
in a variety of cosmological models including $2+1$
gravity\cite{carlip-time}, Gowdy models\cite{viqar-time}
and the Bianchi I model\cite{tate-time} and were found
to give physically meaningful results.  These models were
all exactly solvable, so that expectation values of some
physical quantities could be computed exactly.  We believe
that with path integrals of the form of Eq. (\ref{betaprop}), 
in which gauge
invariance is guaranteed by the construction, it is possible to 
extend
the calculations of physically meaningful quantities in quantum
cosmology to cases such as the Bianchi IX cosmology.

While the Faddeev-Poppov technique
guarantees, if correctly carried out, that the resulting path 
integral is gauge invariant and so represents a physical 
amplitude, exactly which physical amplitude it corresponds
to cannot be defined outside of the context of a consistent
quantization that takes into account the special circumstance
of quantum cosmology.  These include the fact that the universes
described by the Bianchi cosmological model each
live for a finite time, and that the lifetime of each universe
is a function of the initial conditions.  Because the gauge
invariance includes time reparametrization invariance one
must be careful to be sure that one is neither under or
overcounting gauge independent configurations in the path
integral.  Another way to say this is that any quantization of
a cosmological models can only give sensible answers to physical
questions about evolution in time 
if the answers are formulated in a manner that is both
diffeomorphism invariant and takes into account the fact
that any given classical or quantum universe may no longer
exist after a certain span of time.  We take up such issues in the
companion paper.

\section*{ACKNOWLEDGEMENTS}

We would like to thank Abhay Ashtekar, Don Marolf, Nenad
Manojlovic and Jorge Pullin for conversations during the 
course of this work, and also Chris Isham,
Karel Kuchar, Enzo Marinari and Carlo Rovelli for conversations
about the possibility of defining quantum 
cosmological models through
path integrals.  This work has been supported by the
National Science Foundation under grants PHY 9396246 to Syracuse
and Penn State Universities.

\end{document}